\documentclass{aastex631}   
\shorttitle{Homogeneous nucleation of ice in weakly-ionized cold plasma}
\shortauthors{Bellan}
\graphicspath{{./}{figures/}}

\begin{document}

\title{ Mechanism for the efficient homogeneous nucleation of ice in a weakly-ionized, ultra-cold plasma}

\author[00000-0002-0886-8782]{Paul M. Bellan}
\affiliation{Applied Physics and Materials Science, California Institute of Technology\\
1200 E. California Blvd\\
Pasadena, CA 91125, USA}



\begin{abstract}

It is proposed that the rapid observed homogeneous nucleation of ice dust in a
cold, weakly-ionized plasma depends on the formation of hydroxide (OH$^-$)  by fast 
electrons impacting water molecules. These OH$^{-}$ ions attract neutral water molecules because of the
high dipole moment of the water molecules and so hydrates of the form
(OH)$^{-}$(H$_{2}$O)$_{n}$ are formed. The hydrates continuously grow in the cold environment to
become macroscopic ice grains. These ice grains are negatively charged as a
result of electron impact and so continue to attract water molecules. Because
hydroxide  is a negative ion, unlike positive ions it does not suffer
recombination loss from collision with plasma electrons. Recombination with
positive ions is minimal because positive ions are few in number (weak ionization)
and slow-moving as result of being in thermal
equilibrium with the cold background gas.

\end{abstract}



\section{Introduction}

Grains of water ice occur in many terrestrial and extra-terrestrial contexts.
Examples include noctilucent clouds \citep{vasteNoctilucentClouds1993}, comet
tails \citep{protopapaWaterIceDust2014}, Saturn's diffuse rings
\citep{vahidiniaSaturnRingGrains2011}, plumes ejected by Enceladus
\citep{dongCharacteristicsIceGrains2015}, protoplanetary disks
\citep{teradaDetectionWaterIce2007}, and matter orbiting black holes
\citep{moultakaICECUBESCENTER2015}. In order for these grains to nucleate,
there must be pre-existing water vapor or pre-existing hydrogen and oxygen that could make water.
Mechanisms for ice nucleation are categorized as being either  heterogeneous
or homogeneous: heterogeneous nucleation involves ice forming as a cover or
mantle on a pre-existing non-ice solid material such as silicate or
carbon\ whereas\ homogeneous nucleation is the situation where ice forms with
no non-ice substrate. Studies of tropospheric (lower atmosphere) ice formation
have shown that heterogeneous nucleation dominates homogeneous nucleation by
orders of magnitude \citep{pruppacherMicrophysicsCloudsPrecipitation2010}. This
tropospheric domination of heterogeneous nucleation has generally been
presumed to extend to noctilucent clouds (mesosphere)
\citep{gumbelChargedMeteoricSmoke2009} and to extra-terrestrial contexts
\citep{jonesMicronIceBand1984,sekiheterogeneousCondensationInterstellar1983}.
However, these non-tropospheric contexts differ by being ultra-cold (gas
temperature 150 K or less) and weakly ionized.

The paradigm that homogeneous nucleation is negligible compared to
heterogeneous nucleation  requires pre-existence of non-ice solid material. It
is generally presumed that this solid material is silicate
\citep{sekiheterogeneousCondensationInterstellar1983} in the context of
protoplanetary disks.  For noctilucent clouds it is generally presumed
that the solid material is carbon smoke  from ablating micrometeoroids
\citep{gumbelChargedMeteoricSmoke2009}; this solid material is called meteor smoke particle (MSP). However, various observations cast
doubt on this paradigm:

\begin{enumerate}
\item The presumption that extra-terrestrial water ice can exist only as a
coating on silicate cores is clearly invalid for Saturn's rings where in-situ
mass spectrometry of ring ice grains by the Cassini spacecraft Cosmic Dust
Analyzer (CDA) showed that dust grains were composed of either water ice or
silicate \citep{hsuSituCollectionDust2018a}, but not of both.

\item The spectra of Saturn's rings show water ice but no evidence of
silicates \citep{spilkerSaturnRingsThermal2003a}.

\item \citet{potapovDustIceMixing2021a} showed that fits to
certain protoplanetary disk infrared absorption spectra were best explained
using pure water ice rather than a combination of ice and silicates.

\item There is minimal observational evidence for the existence of MSP's at the location where noctilucent clouds form and there is
minimal evidence that micrometeoroids deposit MSP's
\citep{rappModelingMicrophysicsMesospheric2006,hedinMAGICMeteoricSmoke2014}.

\item  \citet{murrayHomogeneousNucleationAmorphous2010a} and
 \citet{zasetskyThermodynamicsHomogeneousNucleation2009a} have
postulated that homogeneous nucleation can occur in noctilucent clouds. An
important feature of the Murray/Jensen and Zasetsky at al. papers is the
realization that at the very low temperatures and pressures of interest,
nucleation likely occurs in a quite different manner from nucleation in the
tropospheric terrestrial atmosphere.

\item An ice dusty plasma experiment by 
\citet{shimizuSynthesisWaterIce2010} at the Max Planck Institute for
Extraterrestrial Physics showed that water ice grains would spontaneously
nucleate in a weakly ionized plasma formed from a mixture of D$_{2}$ and
O$_{2}$ if the electrodes were cooled by liquid nitrogen so that the
background gas would be at a low temperature. Shimuzu et al. modeled their
experiment assuming that the nucleation was heterogeneous (ice nucleated about
non-ice seed particles) but noted that there was no experimental evidence for
the existence of the assumed seed particles.

\item An ice dusty plasma experiment at Caltech
\citep{chaiSpontaneousFormationNonspherical2013,chaiStudyMorphologyGrowth2015,chaiFormationAlignmentElongated2015,marshallIdentificationAccretionGrain2017f}
has shown that ice spontaneously nucleates when a small amount of water vapor
is injected into a weakly ionized plasma if the plasma background gas is cold
(%
$<$%
200 K) and above a critical pressure. Once nucleation has occurred, the
pressure can be lowered at which point the ice grains grow quickly to a large
size within 10-100 seconds. The ice grains are long, slightly dendritic
spindles that are highly charged. The\ system behaves as a classic dusty
plasma with the additional feature that the spindles self-align so as to be
mutually parallel. The experimental parameters are similar to the presumed
parameters of protoplanetary disks and noctilucent clouds.
Specifically, the background gas density of the lab experiment is within one
or two orders of magnitude of these naturally occurring situations, the
background gas temperature is similarly cold, the degree of ionization is
similarly small, and water molecules are similarly present.
These similarities are shown in Table 1. 
Since
protoplanetary disk parameters vary considerably depending on the distance
from the star and the altitude above the midplane, it is quite plausible that
the lab experiment exactly duplicates some specific localized protoplanetary
disk region.

\end{enumerate}

\begin{table}[hbtp] \centering%
\begin{tabular}{llll}
& \textbf{Laboratory} & \textbf{Mesosphere} & \textbf{Protoplanetary} \\ 
& \textbf{experiment} &  & \textbf{disk, midplane} \\ 
\hline\hline
gas temperature  & 190 K & 140 K & 10-100 K \\ 
gas density  & 5$\times $10$^{21}$ m$^{-3}$ &  5$\times $10$^{20}$ m$^{-3}$& 10$^{18}-10^{20}$ m$^{-3}$ \\ 
gas pressure & 10 Pa  & 1 Pa & 0.001-0.1 Pa \\ 
ionization fraction & $10^{-6}$ & $10^{-12}$ & $10^{-10}-10^{-6}$%
\end{tabular}%
\caption{Similarity between three different situations where ice grains form in a weakly ionized plasma having a cold background gas.
Lab experiment from  \citet{chaiSpontaneousFormationNonspherical2013,marshallIdentificationAccretionGrain2017f}; mesophere density from  Table 5 of \citet{rappAbsoluteDensityMeasurements2001}, protoplanetary disk from Figs. 7,9,10, and 12 in \citet{woitkeRadiationThermochemicalModels2009} }%
\label{Situation-comparisons}%
\end{table}%

\section{Classic Nucleation Theory Summarized}

Gas phase ice nucleation is conventionally assumed to occur in accordance with classic
nucleation theory (CNT) which postulates that nucleation is heterogeneous and
results from the interplay between the thermodynamic concepts of Gibbs free
energy and surface tension. CNT was originally formulated for liquid drops but
has been proposed to extrapolate to ice nucleation. CNT assumes that a
new-born liquid drop (in the ice extrapolation, new-born ice grain) is
spherical and has a radius $r$ that relates to the surface tension. This model
was reviewed by \citet{rappModelingMicrophysicsMesospheric2006} and by  \citet{gumbelChargedMeteoricSmoke2009} in the context
of noctilucent cloud ice grains and the process was that of water nucleating
on charged MSP's. A critical assumption in CNT is that
nucleation can be described using bulk matter properties. CNT shows using
considerations of surface tension that there is an energy barrier preventing
nucleation at small radius; this is because the force associated with surface
tension is inversely proportional to the radius of curvature of the surface so
this force becomes infinite as the radius approaches zero.

Gumbel and Megner noted that ionic nucleation could be an alternate to
MSP nucleation. This alternate had been proposed by 
\citet{wittNatureNocitlucentClouds1969} and studied further by 
\citet{sugiyamaIonrecombinationNucleationGrowth1994}; Goldberg and Witt
measured mesospheric ionic content with a sounding rocket. Gumbel and Megner considered
H$^{+}$(H$_{2}$O)$_{n}$ as the embryonic core nucleus and this ionic process
was deemed more favorable than nucleation on a neutral substrate because of
the attraction of the dipole moment of a water molecule to the charge of a
proton. The Gumbel and Megner model was based on CNT and proposed that the
critical nucleus radius at which growth can overcome surface tension is 
$\sim1$ nm. The Gumbel and Megner  model also proposed that the critical
radius becomes zero for the case of ionic nucleation at sufficiently low
temperatures. Although Gumbel and Megner suggested that the ionic scheme could
surpass heterogeneous nucleation, they then dismissed the ionic scheme on the
grounds that ambient electrons would quickly neutralize the proton hydrates so
that this recombination would eliminate the advantages of charge attracting water
molecules. Gumbel and Megner proposed that negatively charged MSP's would instead provide a more plausible nucleus as these would not
suffer recombination via electron collisions and yet would still exploit the
attraction of water molecules to a charged nucleus. 
\citet{hervigContentCompositionMeteoric2012} interpreted optical properties of
noctilucent clouds as indicating that ice grains contained small amounts of
solid material that would be consistent with MSP's and so provide
support for the negatively charged meteor smoke acting as nuclei. However, MSP remains a hypothetical concept as attempts to observe MSP's in an unambiguous way  have so far failed \citep{hedinMAGICMeteoricSmoke2014}. 
As a further example that MSPs remain hypothetical, \citet{studeNovelRocketborneIon2021} interpreted the negative ions observed by a rocket passing through the mesosphere as \textquotedblleft{likely due to negatively charged MSP's}\textquotedblright.

\section{Nucleation from negative ions}

The purpose of this paper is to argue that ice nucleation is indeed ionic as
proposed by Gumbel and Megner but on \textit{negative ions }rather then the
positive ions considered by Gumbel and Megner. A negative ion would attract a
water molecule just like a negatively charged smoke particle and being
negative, would similarly be immune to recombination via electron impact. The
question then is whether the proposed negative ions exist.

There is substantial evidence for the existence of negative ions in various
plasmas but this evidence is generally indirect because direct observation of
negative ions is difficult. The difficulty arises because negative ions have
negligible spectroscopic footprint so their existence cannot be easily
demonstrated by remote spectroscopic observation. Two types of indirect
measurement have been used. The first is to make precise measurements of the
electron and\ the positive ion densities. Since the plasma is expected to be
quasi-neutral, if \ measurements indicate an electron density lower than the
positive ion density, then there must be negative ions to provide the overall
quasi-neutrality. The second method is to transiently detach electrons from
the negative ions using energetic photons and then measure a transient
increase in the electron density using probes or microwaves. The problem with
indirect methods is that their interpretation is complex and requires numerous
assumptions and models. Direct measurement would clearly be preferable. The
most obvious direct measurement method is direct collection of negative ions
but this is impossible for astrophysical situations and difficult for
laboratory situations. On the other hand, negative ions are routinely observed
in the low-altitude atmosphere (troposphere). There is little observational
information at mesospheric altitudes where any depletion of electrons relative
to ions has generally been attributed to dust rather than ions capturing electrons.
The possibility of negative ions in the mesosphere has been dismissed by \citet{baumannInfluencesMeteoricAerosol2016}  on the grounds that atomic oxygen which has significant density in the mesosphere acts as a sink for negative ions; this conclusion was based on laboratory experiments by \citet{fehsenfeldThermalEnergyAssociative1966} investigating the O + O$^- \rightarrow $ O$_2$ + e reaction (the relevance and validity of this conclusion will be addressed in a later paragraph).

Below is a list of several relevant situations showing that plasmas can
contain negative ions. These examples show that plasmas
can contain negative ions of oxygen, hydrogen, hydroxyl, and sulfur hexafluoride.

 \citet{nguyenOperatingRadiofrequencyPlasma2009} measured electron
and ion densities in an argon-water plasma and found that the free electron
density was suppressed compared to that of a pure argon plasma. This was
attributed to electrons attaching to  atoms or molecules such
as hydroxyl and oxygen.

\citet{howlingTimeresolvedMeasurementsHighly1994} observed growth of
highly polymerized negative polysilicon hydride molecules in a silane
(SiH$_{4}$) plasma and suggested that negative ions are the precursors to
particle formation. Because electrons move faster than ions, electrons leave
the bulk plasma faster giving the bulk plasma a positive potential. This
constitutes a potential well for negative ions and so negative ions have near
perfect confinement which provides ample time for negative ions to interact
with other species. The negative ions are presumed to induce a dipole moment
in silane molecules so that there is then an attraction between the negative
ion and the silane. This continues to attract other silane molecules so a
large well-confined negative structure is developed. This process has been
reviewed by \citet{girshickParticleNucleationGrowth2020}.

Using a Wien E$\times$B filter, 
\citet{renaudProbeMeasurementsMolecular2015} observed negative fluorine
atoms in a SF$_{6}$ discharge ion thruster.

\citet{millarNegativeIonsSpace2017a} reviewed negative ions in
astrophysical environments and noted that the ratios of negative hydrogen
atoms and negative hydrocarbon molecules (C$_{4}$H$^{-}$, C$_{6}$H$^{-}$,
etc.) \ to neutrals can be as high as 10\%.

\citet{coatesNegativeIonsEnceladus2010} detected negative ions in
the plume of Enceladus and interpreted these as negative water clusters.

\citet{arshadiHydrationOHO21970} observed gas phase hydrates
of OH$^{-}$, i.e., OH$^{-}$(H$_{2}$O)$_{n}$ in water vapor containing traces
of hydrogen peroxide and in pure water. They also observed gas phase hydrates
of O$_{2}^{-}$, i.e., O$_{2}^{-}$(H$_{2}$O)$_{n}.$ 
\citet{kebarleHydrationNegativeIons1968} observed gas phase hydration of
F$^{-}$, Cl$^{-}$, Br$^{-}$, I$^{-}$, O$_{2}^{-},$ and NO$_{2}^{-}$. \citet{meot-nerFillingSolventShells1986}
also measured production of OH$^{-}$(H$_{2}$O)$_{n}$.  
\citet{newtonInitioStudiesStructures1971} reported ab initio calculations of
the formation \ of OH$^{-}$(H$_{2}$O)$_{n}$.

\citet{sirseMeasurementCF3Densities2017} observed O$^{-}$, F $^{-}%
$, and CF$_{3}^{-}$ in an Ar/O$_{2}$/C$_{4}$F$_{8}$ capacitively coupled rf
discharge using a diagnostic involving laser photo-emission and resonance
shift of a hair-pin probe.

\citet{gregoryInitioMolecularDynamics2019} made ab initio
molecular dynamics studies of the formation of O$_{2}^{-}$(H$_{2}$O)$_{n}$ and
found very stable structures.

\citet{wittNatureNocitlucentClouds1969} considered ion hydrates forming
noctilucent clouds and mainly considered positive ions although mentioned the
possibility of a NO$_{2}^{-}\ $hydrate.
\citet{koppPositiveNegativeIons1992} reported rocket observations of several types of negative ions during a solar eclipse in 1979 at altitudes from 56.2 km to 67.2 km. \citet{studeNovelRocketborneIon2021} reported observations of many types of negative ions at locations inside a polar mesospheric
winter echo layer. 

\citet{jungenFeshbachresonancesDissociativeElectron1979}
presented experimental data showing that electrons with energy exceeding
approximately 4.5 eV would dissociate water to produce hydroxide (negative hydroxyl ion, i.e. OH$^-$);
thus tail electrons in an approximately 2 eV plasma could produce hydroxide.

\citet{shimizuSynthesisWaterIce2010} observed using mass spectrometry an increase in DO (hydroxyl
with deuterium) from background level in their D$_{2}$-H$_{2}$ weakly
ionized plasma when the electrodes were not cooled so no ice dust was formed.
However, they saw a very slight decrease in DO if the electrodes were cooled
so that ice dust was formed.

\citet{fehsenfeldLaboratoryStudiesNegative1974} observed that, in the presence of water vapor, O$^-$ would be depleted and  replaced by OH$^-$ which would then form hydrates of the form OH$^{-}$(H$_{2}$O)$_{n}$. The measurement was done in an O$_2$ background which would be expected to contain atomic oxygen in association with the  O$^-$ production but there apparently was no report of  depletion of OH$^{-}$ by atomic oxygen. This suggests that the conclusion by \citet{baumannInfluencesMeteoricAerosol2016} that all  negative ions would be depleted by atomic oxygen does not occur in the presence of water vapor as the situation becomes more complicated than having only the  O + O$^- \rightarrow $ O$_2$ + e reaction.

\citet{peverallQuantitativeMeasurementsOxygen2020} observed O$^{-}$ in  the presence of  a much larger  atomic oxygen density in an O$_{2}$ discharge. The  O$^{-}$ density was as much as five times larger than the free electron density; this observation further  counters the assertion by \citet{baumannInfluencesMeteoricAerosol2016} that negative ions cannot exist in the presence of atomic oxygen.

Based on the observations listed above, it is proposed here that, as  observed by \citet{fehsenfeldLaboratoryStudiesNegative1974}, negative ions
exist in a weakly ionized plasma containing some combination of water
molecules, hydrogen molecules, and oxygen molecules. \ Since a water molecule
has a very large dipole moment, it will be attracted to a region of strong
electric field; this has been demonstrated by 
\citet{libbrechtCloudChambersCrystal1999} and by 
\citet{bartlettGrowthIceCrystals1963a}. Thus, there will be an attractive force
between a neutral water molecule and a negative ion such as O$_{2}^{-}$ or
OH$^{-}.$ The result of this attraction will be formation of a hydrate such as
the OH$^{-}$(H$_{2}$O)$_{n}$ observed by 
\citet{arshadiHydrationOHO21970} and by \citet{fehsenfeldLaboratoryStudiesNegative1974}. The number $n$ of water molecules in the
hydrate will increase with time because the hydrate will remain charged.
Discharging would require a collision with a positive ion but this will rarely happen because, relative to electrons, ions are very slow
moving. If the temperature is low, the growing hydrate will become an ice
grain rather than a liquid water drop. Thus, the critical ingredients are a
cold, weakly-ionized plasma containing  molecules/atoms with large electron affinities so that
some plasma electrons can attach to the these species to form negative ions to which can  water molecules can attach and create
negative ion hydrates. These ingredients exist in a weakly ionized plasma
containing some combination of water molecules, hydrogen molecules, and oxygen
molecules. It is thus highly likely that the rapid homogeneous nucleation
observed in the Caltech water ice experiment occurs in the sequence: formation
of weakly ionized plasma (free electrons and ions in a sea of neutrals),
dissociation of water into hydroxyl and hydrogen atoms, hydroxide formation, hydrate on hydroxide, growth of hydrate to macroscopic ice
grain. Because of continuous electron and ion bombardment with electron flux
exceeding ion flux, the ice grain will become increasingly negatively charged
in the usual dusty plasma sense \citep{bellanWhyInterstellarIce2020a} as it
becomes larger and so will continue to attract water molecules and grow.

An observed feature of the Caltech ice dusty plasma experiment is that
initiation of nucleation requires operation above a threshold pressure but
once nucleation has begun, it proceeds faster and yields larger particles at
pressures lower than this threshold; this feature is seen no matter what
species of background gas is used. \ For example, 
\citet{marshallIdentificationAccretionGrain2017f} observed that the threshold
for nucleation of ice grains\ in an argon plasma containing a small amount of
water vapor was 600 mT but, once nucleated, the ice grains would grow quickly
at lower pressures. \ The main plasma parameters that change with neutral
pressure are electron temperature and ion density. \citet{boraMeasurementsPlasmaParameters2013} has shown that for a plasma with similar parameters the electron temperature 
decreases with increased pressure  while the density has a non-monotonic
behavior, first increasing and then decreasing (see Fig.10 in  \citet{boraMeasurementsPlasmaParameters2013}). The production of negative
ions is likely strongly affected by electron temperature because the electron
affinity is comparable to electron temperatures whereas the effect of electron
density is expected to be less because, unlike electron affinity, there is no
resonance-like condition.

A related observed feature of the Caltech ice dusty plasma is that nucleation
does not occur if there is too much rf power. Specifically, 
\citet{chaiStudyMorphologyGrowth2015} reported that rf power must not be higher
than a certain value (0.5 W) but then once nucleation occurred the rf power
could be increased and then the ice grains would grow to a large size.
Operation above a critical rf power likely corresponds to operating above a
critical electron temperature suggesting that nucleation is inhibited if the
temperature of bulk electrons exceeds some critical value; Fig. 6 in  \citet{boraMeasurementsPlasmaParameters2013} shows that the electron temperature increases with rf power.

This suggests that the bulk electron temperature should not be so high as to
knock electrons off from negative ions. The most likely candidates for
forming negative ions in a plasma containing water molecules are atomic
hydrogen, atomic or molecular oxygen, and hydroxyl radicals; electron affinities 
$E_{a}$ for these species are listed in Table 2.
\begin{table}[tbp] \centering
\begin{tabular}
[c]{|l|l|}\hline
\textbf{ species} & \textbf{electron affinity energy }$E_{a}$\textbf{
(eV)}\\\hline
O & 1.43\\
O$_{2}$ & 0.45\\
H & 0.76\\
OH & 1.82\\\hline
\end{tabular}
\caption{Electron affinities of atoms and molecules obtainable from water}\label{TableKey}%
\end{table}%

Table 2 suggests that the most likely candidate for dominant negative ion in a
water-containing capacitive discharge plasma is  hydroxide (OH$^-$)  as the
electron temperature in such a plasma is $\sim$2 eV
\citep{boraMeasurementsPlasmaParameters2013}. Plasma electrons at 2 eV would
thus have more than enough energy to knock electrons off from O$_{2}$
molecules or H atoms, but 2 eV electron energy would be marginal or
insufficient for detaching bound electrons from O atoms or OH radicals.
\ \ While O has an electron affinity similar to OH, as will be seen below,
generation of O from water requires more energy than generation of OH so the
O\ path is less favorable. As support for the contention that OH$^{-}$ is
likely the predominant negative ion, it is noted that 
\citet{nguyenOperatingRadiofrequencyPlasma2009} saw evidence of strong OH
peaks in both the optical and mass spectra of a water plasma but only a small
O$_{2}$ peak in the mass spectrum and an even smaller O\ peak. \citet{sturmFirstResultsHerschel2010} observed simultaneous existence of
\ H$_{2}$O, O, and OH in a protoplanetary disk so the necessary ingredients
for forming negative ion hydrates exist in a protoplanetary disk, namely cold
background gas, weak ionization, H$_{2}$O, O, and OH. It is thus expected
based on the energy arguments from the lab plasma context that in a
protoplanetary disk hydroxide would also be the most likely
negative ion for forming hydrates.

\citet{cottinEtudeIonsProduits1959} provided extensive measurements of
how the production of positive and negative ions from electron bombardment of
water vapor depends on electron kinetic energy $E.$ These measurements showed
that ion production peaks at specific values of $E$; more recent data has been
provided by  \citet{fedorFragmentationTransientWater2006}. Table 3
lists Cottin's results for the lowest electron energy $E$ that \ produce
negative ions from water molecules and also the kinetic energies denoted as $E_{p}$ at which this
negative ion production is at a peak. The energies reported by Cottin for
production of positive ions by electron bombardment of water molecules are
$\geq$12.6 V so in a plasma where $T_{e}\ll12.6$ eV \ negative ions should be
the dominant type of ion produced by electron bombardment of water molecules. \ %

\begin{table}[tbp] \centering
\begin{tabular}
[c]{|l|l|l|l|}\hline
\textbf{Ion } & \textbf{Appearance (eV)} & \textbf{Peak production }$E_{p}%
$\textbf{ (eV)} & \textbf{Formation mechanism}\\\hline
H$^{-}$ & 4.8 & 6.0, 8.0 (weak) & H$_{2}$O + e $\rightarrow$ H$^{-}$ + OH\\
OH$^{-}$ & 4.7 & 6.1, 8.5, 11.2 & H$^{-}+$ H$_{2}$O $\rightarrow$ OH$^{-}+$
H$_{2}$\\
O$^{-}$ & 7.4 & 9.2, 11.2 & H$_{2}$O + e $\rightarrow$O$^{-}+2$H (or H$_{2}%
$)\\\hline
\end{tabular}
\caption{Electron energy in eV  at which negative ions first appears when water vapor is bombarded with electrons, electron energy at which negative ion yield has peak and presumed reaction. Data is from Cottin and uncertainties in potential are typically +/- 0.2 volts}\label{TableKey copy(1)}%
\end{table}%

\citet{cottinEtudeIonsProduits1959} observed that the production of
OH$^{-}$ was proportional to the square of the water vapor pressure whereas
the production of H$^{-}$ and O$^{-}$ was linearly proportional and concluded
from this observation that OH$^{-}$ production\ is a two-step process where
the first step is either%
\begin{equation}
\text{H}_{2}\text{O + e }\rightarrow\text{H}^{-}+\text{OH} \label{1}%
\end{equation}
or%
\begin{equation}
\text{H}_{2}\text{O + e }\rightarrow\text{O}^{-}+\text{2H (or H}_{2}\text{)}
\label{2}%
\end{equation}
that would then be followed by a second step of either%
\begin{equation}
\text{H}^{-}+\text{H}_{2}\text{O }\rightarrow\text{OH}^{-}\text{ + H}_{2}
\label{3}%
\end{equation}
or%
\begin{equation}
\text{O}^{-}+\text{H}_{2}\text{O }\rightarrow\text{OH}^{-}\text{ + OH.}
\label{4}%
\end{equation}
The need for two steps, each involving H$_{2}$O, would lead to the quadratic
dependence on water vapor pressure. \citet{fedorFragmentationTransientWater2006} observed a linear dependence of
OH$^{-}$ formation \ on water vapor pressure but the Fedor et al. measurements
\ were made at a much lower water vapor pressure so the quadratic path might
have become insignificant relative to some linear single step path. Also,
Fedor et al. noted theoretical arguments against a linear dependence. \ The
path involving H$^{-}$ in the second step (i.e., Eq.\ref{3}) seems more likely
than the path involving O$^{-}$ (i.e., Eq.\ref{4}) because, as listed in Table
3, the electron energy required to form O$^{-}$ is substantially higher than
that required to form H$^{-}$. Cottin's plot of OH$^{-}$ production as a
function of electron energy shows three maxima where the lowest of the three
peaks (6 volts) corresponds to a peak for the production of H$^{-}$ and the
highest of the three peaks (11.2 V) corresponds to a peak for the production
of O$^{-}$.

It is proposed here that in order to produce negative hydrates in a weakly
ionized plasma containing water vapor, a low $T_{e}$ is desirable to avoid
knocking electrons off from electronegative ions while a high $T_{e}$ is
desirable to create these electronegative ions. Because these requirements are
conflicting, the optimum $T_{e}$ should have a value intermediate between the
affinity energy of a negative ion and the lowest electron energy peak for
creating the negative ion. For hydroxyde, the lowest electron
energy peak is 6 eV. 

By expressing the  electron Maxwell-Boltzmann distribution in terms of energy,
the probability that an electron has a kinetic energy between $E$ and $E+dE$
is%
\begin{equation}
f(E)dE=2\sqrt{\frac{E}{\pi}}\frac{1}{\left(  T_{e}\right)  ^{3/2}}\exp\left(
-\frac{E}{T_{e}}\right)  dE\label{f}%
\end{equation}
where the electron kinetic energy $E$ and the electron temperature $T_{e}$ are both measured in electron volts.
Electrons  that have kinetic energy $E$ exceeding the affinity energy $E_{a}$
of an electron attached to a negative ion will detach the bound electron, in a manner similar to  an energetic electron ionizing a neutral atom or molecule. This is analogous to laser
photo-detachment \citep{sirseMeasurementCF3Densities2017} where a photon with
energy exceeding $E_{a}$ detaches the bound electron  from a negative ion. The
flux $F_{d}$ of ``detaching" electrons having  $E>E_{a}$ is
\begin{equation}
 F_{d}   =n_{e}\int_{E_{a}}^{\infty}\left(  \frac{2E}{m_{e}}\right)
^{1/2}f(E)dE\nonumber\\
 =n_{e}\left(  \frac{8T_{e}}{\pi m}\right)  ^{1/2}\left(  \frac{E_{a}}%
{T_{e}}+1\right)  \exp\left(  -\frac{E_{a}}{T_{e}}\right)  .\label{F}%
\end{equation}

\bigskip

The flux of attaching electrons, i.e., electrons having energy at the  energy
$E_{p}$ as listed in the third column of Table 3,  is found by multiplying
Eq.\ref{f} by the electron velocity $v\propto\sqrt{E_{p}}$ and density $n_{e}$
to obtain
\begin{equation}
F_{a}=\lambda n_{e}\frac{E_{p}}{\left(  T_{e}\right)  ^{3/2}}\exp\left(
-\frac{E_{p}}{T_{e}}\right)  \label{Fa}%
\end{equation}
where  $\lambda$ is a coefficient proportional to the width of the peak. 

Thus $F_{a}$ characterizes the rate at which negative ions are created while
$F_{d}$ describes the rate at which they are destroyed. The ratio of these two
fluxes consequently gives  a rough figure of merit for the value of $T_{e}$ that maximizes creation of negative ions.
This ratio is
\begin{equation}
\frac{F_{a}}{F_{d}}=\mu\frac{E_{p}}{\left(  T_{e}\right)  ^{2}}\frac
{\exp\left(  \frac{E_{a}-E_{p}}{T_{e}}\right)  }{\left(  \frac{E_{a}}{T_{e}%
}+1\right)  }\label{ratio}%
\end{equation}
where $\mu$ is a constant incorporating the factors that do not involve
$T_{e}.$ 

\bigskip%
\begin{figure}[ptb]%
\centering
\includegraphics[
height=5.3895in,
width=6.7308in
]%
{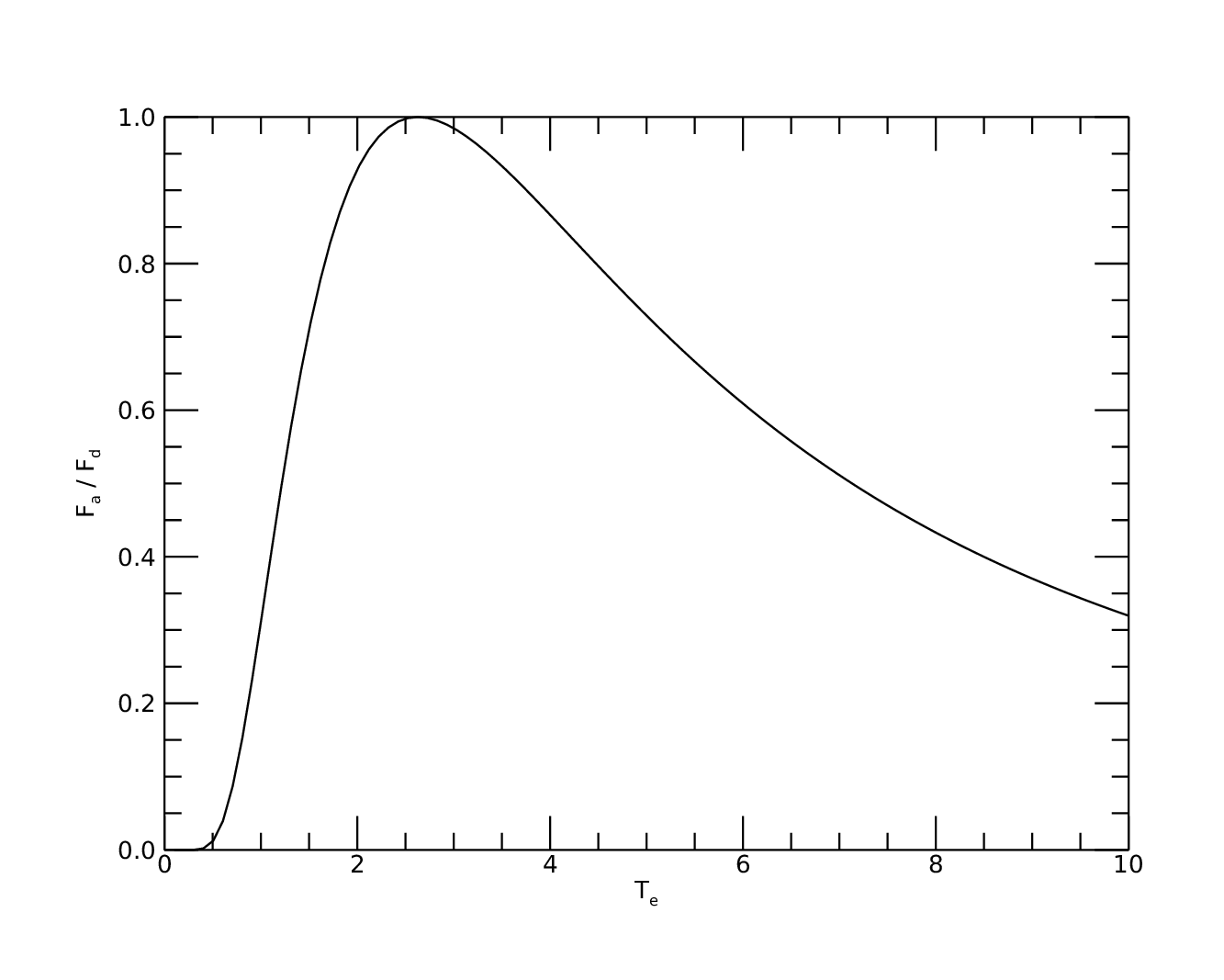}%
\caption{Plot of $F_{a}/F_{d}$ normalized to its maximum value versus $T_{e}$
for hydroxyde formation using electron affinity energy
$E_{a}=1.82$ eV and peak electron energy \citep{cottinEtudeIonsProduits1959} for formation $E_{p}=6$ eV.}%
\end{figure}

Figure 1 plots $\ F_{a}/F_{d}$ for the hydroxyde situation where $E_{a}=1.82$
eV and $E_{p}=6$ eV. This plot  peaks at $T_{e}=2.65$ eV which is of the
order of the $T_{e}$ presumed for the Caltech ice dusty plasma. This ratio
$F_{a}/F_{d}$ is a rough figure of merit because it does not  account
for\ additional issues such as stronger detachment of bound electrons by
incident  electrons with $E\gg E_{a}$ compared to  those with $E\gtrsim E_{a}%
$ and the possibility of a non-Maxellian two temperature electron velocity distribution such as reported by \citet{godyakAbnormallyLowElectron1990}. The peak in Fig.1 at a low but finite electron temperature  is consistent
with  the observation that nucleation in a weakly-ionized plasma occurs at low
rf power \citep{chaiStudyMorphologyGrowth2015} and at high background gas
pressure \citep{marshallIdentificationAccretionGrain2017f}, since both of these
regimes imply low $T_{e}.$ \citet{shimizuSynthesisWaterIce2010} 
observed (Fig. 5 in their paper) an increase of DO upon formation of a
weakly ionized deuterium-oxygen plasma when the electrodes were not cooled (no
ice nucleation, Fig. 5b) but a slight decrease of DO when the electrodes were
cooled (ice dust nucleation seen, Fig. 5a). This observation  is consistent
with DO being absorbed into negative ion hydrates DO$^{-}$(D$_{2}$O)$_{n}$ as the
initial step in homogeneous ice nucleation in a weakly ionized cold plasma
containing D$_{2}$O molecules. Shimuzu et al.'s observation in their Fig.5c
that DO increased when the cooling was removed is consistent with DO no longer
being absorbed when ice nucleation stops. This is circumstantial evidence
because a conventional mass spectrometer does not observe negative ions since
its internal acceleration is arranged for positive ions. Shimizu et al. were
thus likely measuring neutral DO  but this  could be considered a proxy for
DO$^{-}$.

\bigskip

\bigskip
\bigskip
\section{Critique of CNT}

The plausibility of this negative ion nucleation hypothesis is bolstered by
noting that certain aspects of CNT involve using a theory outside its range of
validity. CNT is based on thermodynamics but the essential assumption
underlying thermodynamics is inappropriate for a dusty plasma because the
electrons are much hotter than the ions and neutrals (e.g., electron
temperature is of the order of $10^{4}$ K whereas ions and neutral temperature
are of the order of $10-10^{2}$ K). \ The essential assumption of
thermodynamics, namely that all constituents are at the same temperature, is
thus clearly violated.  The Gibbs free
energy, an important quantity in CNT, is undefined in a weakly ionized plasma
because Gibbs free energy is defined for fixed pressure and fixed temperature.
Because a weakly ionized plasma cannot be characterized by a single
temperature, the Gibbs free energy is not meaningful in a weakly ionized plasma.

A related aspect that casts doubt on CNT is that application of CNT to ice
nucleation presumes that the surface tension of liquid water can be
extrapolated to apply to ice, but extrapolating surface tension from liquid to
solid is debatable as noted by 
\citet{makkonenMisinterpretationShuttleworthEquation2012} who showed that
solids do not have surface tension in the same sense as liquids. The surface
of ice in terrestrial conditions is known to be coated by a quasi-liquid layer
(QLL) that is a few molecules thick. At first sight, it seems that a QLL could
conceivably provide a form of surface tension and so provide a basis for CNT.
However,  \citet{slaterSurfacePremeltingWater2019} showed
that the QLL vanishes at temperatures
$<$
240 K so ice nucleation models for ordinary terrestrial conditions (i.e.,
temperature exceeding 240 K) are inapplicable to the much colder astrophysical
regime (i.e., $T\ll240$ K). Concern about invoking surface tension in CNT even
for liquid drops was noted by  \citet{tolmanEffectDropletSize1949} who
stated \textquotedblleft\textit{As we consider smaller and smaller droplets of
liquid phase, however, the concept of surface tension and the previously
satisfactory thermodynamic methods seem less and less appropriate. Indeed, it
will ultimately seem more satisfactory to continue the investigation using the
concept of forces exerted by individual molecules and the more detailed
methods of molecular mechanics.}\textquotedblright\ Homogeneous nucleation
must begin by having two molecules bind together but at this stage no surface
exists so thermodynamic arguments involving surface tension are inappropriate.
Thermodynamics implicitly presumes the presence of a large number of
molecules, not two.

It should be noted that \citet{murrayHomogeneousNucleationAmorphous2010a}  have previously advocated the possibility of homogeneous nucleation in the mesosphere, but the mechanism they proposed involved neutral water molecules  rather than the  negative ions proposed here. The negative ion mechanism  should be  more important because it results in a stronger bond. The first step in homogeneous nucleation involves two molecules sticking together. The effectiveness of the  homogeneous water-water process proposed by \citet{murrayHomogeneousNucleationAmorphous2010a} can thus be compared to the hydroxide hydrate process proposed here by comparing how strongly the initial reacting two molecules stick together.  This is done by comparing the respective enthalpy heats of reaction $\Delta H_{r}^{0}$ of the Murray-Jensen mechanism and of the hydroxide hydrate mechanism proposed here. The Murray-Jensen homogeneous nucleation has the reaction 
\begin{equation}
\text{H}_{2}\text{O }+\text{ H}_{2}\text{O}\rightarrow \text{(H}_{2}\text{O)}%
_{2}\text{; }\Delta H_{r}^{0}=-9\text{ kJ/mole}  \label{Murray-Jensen}
\end{equation}
whereas the process proposed here has  the reaction
\begin{equation}
\text{OH}^{-}\text{ }+\text{ H}_{2}\text{O}\rightarrow \text{OH}^{-}\text{(H}%
_{2}\text{O); }\Delta H_{r}^{0}=-115\text{ kJ/mole}.
\label{hydroxide hydrate}
\end{equation}
The reaction enthalpies are calculated using the Argonne National Lab Active Thermochemical Tables \citep{ruscicActiveThermochemicalTables2021}.
The hydroxide-water reaction has a much more negative enthalpy heat of reaction which indicates that  OH$^{-}$(H$_{2}$O)  is a more stable state than  two attached water molecules and so should be more important.  This is in accord with the observation by \citet{fehsenfeldLaboratoryStudiesNegative1974} that strong bonding occurs when OH$^-$ associates with H$_2$O.

Stability of  hydroxide hydride  with respect to destruction by atomic oxygen 
can be checked by considering 
\begin{equation}
\text{O}+2\text{[(OH-)(H}_{2}\text{O)]}\rightarrow 3\text{H}_{2}\text{O}+%
\text{O}_{2}+2e  \label{O and hydroxide hydrate}
\end{equation}%
where the stoichiometry has been arranged so the products are standard
states in which case no further spontaneous reactions should occur (water
does not burn). The reaction enthalpy here is $\Delta H_{r}^{0}=+9.4$
kJ/mole; thus this reaction is endothermic and  would not occur spontaneously
(see discussion in \citet{fehsenfeldThermalEnergyAssociative1966} where it was noted that reactions with \ $%
\Delta H_{r}^{0}>0$ were not observed). This means that hydroxide hydride is a stable state and will not be destroyed by interaction with atomic oxygen.
This is in contrast to the reaction 
\begin{equation}
\text{O}+\text{O}^{-}\rightarrow \text{O}_{2}+e  \label{O and O-}
\end{equation}%
which has a reaction enthalpy $\Delta H_{r}^{0}=-357$ kJ/mole and so is
spontaneous, implying that oxygen anion will not endure in the presence of atomic oxygen unless oxygen anion is being continuously replenished.

In summary, the standard presumption that heterogeneous nucleation dominates
homogeneous nucleation does not apply to the situation of a weakly ionized
plasma consisting of a cold background gas with negative atoms or
molecules. Since OH and H$_{2}$O commonly exist in various astrophysical situations, in the mesosphere, and exist  in the ice dust lab experiments \citep{chaiFormationAlignmentElongated2015,shimizuSynthesisWaterIce2010}, and
since OH\ has a large electron affinity, homogeneous nucleation of ice via
hydrates based on OH$^{-}$ is likely to  be operative in these situations. 
The OH$^{-}$ density does not have to be large since each OH$^{-}$ ion forms the
nucleus for an ice grain that would have many attached electrons. The  
required  OH$^{-}$ density is thus so low that it would be difficult to prove that the OH$^{-}$
density is inadequate. 

\ The semi-quantitative \ arguments presented here provide guidance for future
more precise analysis and measurements to determine which negative ion is
indeed dominant and how the process scales to different parameter regimes.
Motivated by these arguments, a diagnostic extending to a water-ice dusty
plasma the methods used in 
\citet{howlingTimeresolvedMeasurementsHighly1994} is being constructed. This
diagnostic will search for the negative ions predicted to exist as a precursor
to the homogeneous nucleation of water ice and will determine how the negative
ions depend on plasma parameters such as ion density, electron temperature,
and background gas species and whether negative ion formation is the initial
step to ice grain nucleation in a cold weakly-ionized plasma containing water vapor.

Acknowledgements: Supported by the NSF/DOE Partnership in Basic Plasma Science
and Engineering\ via USDOE\ Award DE-SC0020079.

\bibliographystyle{aasjournal}
\bibliography{2021-negative-ion-paperv2}

\begin{thebibliography}{}
\expandafter\ifx\csname natexlab\endcsname\relax\def\natexlab#1{#1}\fi
\providecommand{\url}[1]{\href{#1}{#1}}
\providecommand{\dodoi}[1]{doi:~\href{http://doi.org/#1}{\nolinkurl{#1}}}
\providecommand{\doeprint}[1]{\href{http://ascl.net/#1}{\nolinkurl{http://ascl.net/#1}}}
\providecommand{\doarXiv}[1]{\href{https://arxiv.org/abs/#1}{\nolinkurl{https://arxiv.org/abs/#1}}}

\bibitem[{Arshadi \& Kebarle(1970)}]{arshadiHydrationOHO21970}
Arshadi, M., \& Kebarle, P. 1970, The Journal of Physical Chemistry, 74, 1483,
  \dodoi{10.1021/j100702a015}

\bibitem[{Bartlett {et~al.}(1963)Bartlett, Heuval, \&
  Mason}]{bartlettGrowthIceCrystals1963a}
Bartlett, J.~T., Heuval, A.~P., \& Mason, B.~J. 1963, Zeitschrift fur
  angewandte Mathematik und Physik ZAMP, 14, 599, \dodoi{10.1007/BF01601267}

\bibitem[{Baumann(2016)}]{baumannInfluencesMeteoricAerosol2016}
Baumann, C. 2016, {PhD}, Ludwig Maximilians University, Muncih

\bibitem[{Bellan(2020)}]{bellanWhyInterstellarIce2020a}
Bellan, P.~M. 2020, The Astrophysical Journal, 905, 96,
  \dodoi{10.3847/1538-4357/abc55b}

\bibitem[{Bora {et~al.}(2013)Bora, Bhuyan, Favre, Wyndham, Chuaqui, \&
  Wong}]{boraMeasurementsPlasmaParameters2013}
Bora, B., Bhuyan, H., Favre, M., {et~al.} 2013, Current Applied Physics, 13,
  1448, \dodoi{10.1016/j.cap.2013.04.024}

\bibitem[{Chai \& Bellan(2013)}]{chaiSpontaneousFormationNonspherical2013}
Chai, K.-B., \& Bellan, P.~M. 2013, Geophysical Research Letters, 40, 6258,
  \dodoi{10.1002/2013GL058268}

\bibitem[{Chai \& Bellan(2015{\natexlab{a}})}]{chaiStudyMorphologyGrowth2015}
---. 2015{\natexlab{a}}, Journal of Atmospheric and Solar-Terrestrial Physics,
  127, 83, \dodoi{10.1016/j.jastp.2014.07.012}

\bibitem[{Chai \&
  Bellan(2015{\natexlab{b}})}]{chaiFormationAlignmentElongated2015}
---. 2015{\natexlab{b}}, Astrophysical Journal, 802, 112,
  \dodoi{10.1088/0004-637X/802/2/112}

\bibitem[{Coates {et~al.}(2010)Coates, Jones, Lewis, Wellbrock, Young, Crary,
  Johnson, Cassidy, \& Hill}]{coatesNegativeIonsEnceladus2010}
Coates, A., Jones, G., Lewis, G., {et~al.} 2010, Icarus, 206, 618,
  \dodoi{10.1016/j.icarus.2009.07.013}

\bibitem[{Cottin(1959)}]{cottinEtudeIonsProduits1959}
Cottin, M. 1959, Journal de Chimie Physique, 56, 1024,
  \dodoi{10.1051/jcp/1959561024}

\bibitem[{Dong {et~al.}(2015)Dong, Hill, \&
  Ye}]{dongCharacteristicsIceGrains2015}
Dong, Y., Hill, T.~W., \& Ye, S.-Y. 2015, Journal of Geophysical Research:
  Space Physics, 120, 915, \dodoi{10.1002/2014JA020288}

\bibitem[{Fedor {et~al.}(2006)Fedor, Cicman, Coupier, Feil, Winkler, Gluch,
  Husarik, Jaksch, Farizon, Mason, Scheier, \&
  Mark}]{fedorFragmentationTransientWater2006}
Fedor, J., Cicman, P., Coupier, B., {et~al.} 2006, Journal of Physics B:
  Atomic, Molecular and Optical Physics, 39, 3935,
  \dodoi{10.1088/0953-4075/39/18/022}

\bibitem[{Fehsenfeld \&
  Ferguson(1974)}]{fehsenfeldLaboratoryStudiesNegative1974}
Fehsenfeld, F.~C., \& Ferguson, E.~E. 1974, The Journal of Chemical Physics,
  61, 3181, \dodoi{10.1063/1.1682474}

\bibitem[{Fehsenfeld {et~al.}(1966)Fehsenfeld, Ferguson, \&
  Schmeltekopf}]{fehsenfeldThermalEnergyAssociative1966}
Fehsenfeld, F.~C., Ferguson, E.~E., \& Schmeltekopf, A.~L. 1966, The Journal of
  Chemical Physics, 45, 1844, \dodoi{10.1063/1.1727844}

\bibitem[{Girshick(2020)}]{girshickParticleNucleationGrowth2020}
Girshick, S.~L. 2020, Journal of Vacuum Science \& Technology A, 38, 011001,
  \dodoi{10.1116/1.5136337}

\bibitem[{Godyak \& Piejak(1990)}]{godyakAbnormallyLowElectron1990}
Godyak, V.~A., \& Piejak, R.~B. 1990, Physical Review Letters, 65, 996,
  \dodoi{10.1103/PhysRevLett.65.996}

\bibitem[{Gregory {et~al.}(2019)Gregory, Reveles, Bly, \&
  Luong}]{gregoryInitioMolecularDynamics2019}
Gregory, N., Reveles, J.~U., Bly, J., \& Luong, T. 2019, The Journal of
  Physical Chemistry A, 123, 7528, \dodoi{10.1021/acs.jpca.9b04510}

\bibitem[{Gumbel \& Megner(2009)}]{gumbelChargedMeteoricSmoke2009}
Gumbel, J., \& Megner, L. 2009, Journal of Atmospheric and Solar-Terrestrial
  Physics, 71, 1225, \dodoi{10.1016/j.jastp.2009.04.012}

\bibitem[{Hedin {et~al.}(2014)Hedin, Giovane, Waldemarsson, Gumbel, Blum,
  Stroud, Marlin, Moser, Siskind, Jansson, Saunders, Summers, Reissaus,
  Stegman, Plane, \& Horányi}]{hedinMAGICMeteoricSmoke2014}
Hedin, J., Giovane, F., Waldemarsson, T., {et~al.} 2014, Journal of Atmospheric
  and Solar-Terrestrial Physics, 118, 127, \dodoi{10.1016/j.jastp.2014.03.003}

\bibitem[{Hervig {et~al.}(2012)Hervig, Deaver, Bardeen, Russell, Bailey, \&
  Gordley}]{hervigContentCompositionMeteoric2012}
Hervig, M.~E., Deaver, L.~E., Bardeen, C.~G., {et~al.} 2012, Journal of
  Atmospheric and Solar-Terrestrial Physics, 84-85, 1,
  \dodoi{10.1016/j.jastp.2012.04.005}

\bibitem[{Howling {et~al.}(1994)Howling, Sansonnens, Dorier, \&
  Hollenstein}]{howlingTimeresolvedMeasurementsHighly1994}
Howling, A.~A., Sansonnens, L., Dorier, J.-L., \& Hollenstein, C. 1994, Journal
  of Applied Physics, 75, 1340, \dodoi{10.1063/1.356413}

\bibitem[{Hsu {et~al.}(2018)Hsu, Schmidt, Kempf, Postberg,
  Moragas-Klostermeyer, Seiss, Hoffmann, Burton, Ye, Kurth, Horanyi, Khawaja,
  Spahn, Schirdewahn, O'Donoghue, Moore, Cuzzi, Jones, \&
  Srama}]{hsuSituCollectionDust2018a}
Hsu, H.-W., Schmidt, J., Kempf, S., {et~al.} 2018, Science, 362, eaat3185,
  \dodoi{10.1126/science.aat3185}

\bibitem[{Jones \& Williams(1984)}]{jonesMicronIceBand1984}
Jones, A.~P., \& Williams, D.~A. 1984, Monthly Notices of the Royal
  Astronomical Society, 209, 955, \dodoi{10.1093/mnras/209.4.955}

\bibitem[{Jungen {et~al.}(1979)Jungen, Vogt, \&
  Staemmler}]{jungenFeshbachresonancesDissociativeElectron1979}
Jungen, M., Vogt, J., \& Staemmler, V. 1979, Chemical Physics, 37, 49,
  \dodoi{10.1016/0301-0104(79)80005-1}

\bibitem[{Kebarle {et~al.}(1968)Kebarle, Arshadi, \&
  Scarborough}]{kebarleHydrationNegativeIons1968}
Kebarle, P., Arshadi, M., \& Scarborough, J. 1968, The Journal of Chemical
  Physics, 49, 817, \dodoi{10.1063/1.1670145}

\bibitem[{Kopp(1992)}]{koppPositiveNegativeIons1992}
Kopp, E. 1992, Advances in Space Research, 12, 325,
  \dodoi{10.1016/0273-1177(92)90488-J}

\bibitem[{Libbrecht \& Tanusheva(1999)}]{libbrechtCloudChambersCrystal1999}
Libbrecht, K.~G., \& Tanusheva, V.~M. 1999, Physical Review E, 59, 3253,
  \dodoi{10.1103/PhysRevE.59.3253}

\bibitem[{Makkonen(2012)}]{makkonenMisinterpretationShuttleworthEquation2012}
Makkonen, L. 2012, Scripta Materialia, 66, 627,
  \dodoi{10.1016/j.scriptamat.2012.01.055}

\bibitem[{Marshall {et~al.}(2017)Marshall, Chai, \&
  Bellan}]{marshallIdentificationAccretionGrain2017f}
Marshall, R.~S., Chai, K.-B., \& Bellan, P.~M. 2017, Astrophysical Journal,
  837, 56, \dodoi{10.3847/1538-4357/aa5d11}

\bibitem[{Meot-Ner \& Speller(1986)}]{meot-nerFillingSolventShells1986}
Meot-Ner, M., \& Speller, C.~V. 1986, The Journal of Physical Chemistry, 90,
  6616, \dodoi{10.1021/j100283a006}

\bibitem[{Millar {et~al.}(2017)Millar, Walsh, \&
  Field}]{millarNegativeIonsSpace2017a}
Millar, T.~J., Walsh, C., \& Field, T.~A. 2017, Chemical Reviews, 117, 1765,
  \dodoi{10.1021/acs.chemrev.6b00480}

\bibitem[{Moultaka {et~al.}(2015)Moultaka, Eckart, \&
  Muzic}]{moultakaICECUBESCENTER2015}
Moultaka, J., Eckart, A., \& Muzic, K. 2015, The Astrophysical Journal, 806,
  202, \dodoi{10.1088/0004-637X/806/2/202}

\bibitem[{Murray \& Jensen(2010)}]{murrayHomogeneousNucleationAmorphous2010a}
Murray, B.~J., \& Jensen, E.~J. 2010, Journal of Atmospheric and
  Solar-Terrestrial Physics, 72, 51, \dodoi{10.1016/j.jastp.2009.10.007}

\bibitem[{Newton \& Ehrenson(1971)}]{newtonInitioStudiesStructures1971}
Newton, M.~D., \& Ehrenson, S. 1971, Journal of the American Chemical Society,
  93, 4971, \dodoi{10.1021/ja00749a001}

\bibitem[{Nguyen {et~al.}(2009)Nguyen, Foster, \&
  Gallimore}]{nguyenOperatingRadiofrequencyPlasma2009}
Nguyen, S. V.~T., Foster, J.~E., \& Gallimore, A.~D. 2009, Review of Scientific
  Instruments, 80, 083503, \dodoi{10.1063/1.3202250}

\bibitem[{Peverall {et~al.}(2020)Peverall, Rogers, \&
  Ritchie}]{peverallQuantitativeMeasurementsOxygen2020}
Peverall, R., Rogers, S. D.~A., \& Ritchie, G. A.~D. 2020, Plasma Sources
  Science and Technology, 29, 045004, \dodoi{10.1088/1361-6595/ab7840}

\bibitem[{Potapov {et~al.}(2021)Potapov, Bouwman, Jäger, \&
  Henning}]{potapovDustIceMixing2021a}
Potapov, A., Bouwman, J., Jäger, C., \& Henning, T. 2021, Nature Astronomy, 5,
  78, \dodoi{10.1038/s41550-020-01214-x}

\bibitem[{Protopapa {et~al.}(2014)Protopapa, Sunshine, Feaga, Kelley, A'Hearn,
  Farnham, Groussin, Besse, Merlin, \& Li}]{protopapaWaterIceDust2014}
Protopapa, S., Sunshine, J.~M., Feaga, L.~M., {et~al.} 2014, Icarus, 238, 191,
  \dodoi{10.1016/j.icarus.2014.04.008}

\bibitem[{Pruppacher \&
  Klett(2010)}]{pruppacherMicrophysicsCloudsPrecipitation2010}
Pruppacher, H., \& Klett, J. 2010, Atmospheric and {Oceanographic} {Sciences}
  {Library}, Vol.~18, Microphysics of {Clouds} and {Precipitation} (Dordrecht:
  Springer Netherlands), \dodoi{10.1007/978-0-306-48100-0}

\bibitem[{Rapp {et~al.}(2001)Rapp, Gumbel, \&
  Lübken}]{rappAbsoluteDensityMeasurements2001}
Rapp, M., Gumbel, J., \& Lübken, F.-J. 2001, Annales Geophysicae, 19, 571,
  \dodoi{10.5194/angeo-19-571-2001}

\bibitem[{Rapp \& Thomas(2006)}]{rappModelingMicrophysicsMesospheric2006}
Rapp, M., \& Thomas, G.~E. 2006, Journal of Atmospheric and Solar-Terrestrial
  Physics, 68, 715, \dodoi{10.1016/j.jastp.2005.10.015}

\bibitem[{Renaud {et~al.}(2015)Renaud, Gerst, Mazouffre, \&
  Aanesland}]{renaudProbeMeasurementsMolecular2015}
Renaud, D., Gerst, D., Mazouffre, S., \& Aanesland, A. 2015, Review of
  Scientific Instruments, 86, 123507, \dodoi{10.1063/1.4937604}

\bibitem[{Ruscic \& Bross(2021)}]{ruscicActiveThermochemicalTables2021}
Ruscic, B., \& Bross, D.~H. 2021, Active {Thermochemical} {Tables} ({ATcT}),
  Argonne National Lab.
\newblock \url{ATcT.anl.gov}

\bibitem[{Seki \&
  Hasegawa(1983)}]{sekiheterogeneousCondensationInterstellar1983}
Seki, J., \& Hasegawa, H. 1983, Astrophysics and Space Science, 94, 177,
  \dodoi{10.1007/BF00651770}

\bibitem[{Shimizu {et~al.}(2010)Shimizu, Klumov, Shimizu, Rothermel, Havnes,
  Thomas, \& Morfill}]{shimizuSynthesisWaterIce2010}
Shimizu, S., Klumov, B., Shimizu, T., {et~al.} 2010, Journal of Geophysical
  Research-Atmospheres, 115, D18205, \dodoi{10.1029/2009JD013375}

\bibitem[{Sirse {et~al.}(2017)Sirse, Tsutsumi, Sekine, Hori, \&
  Ellingboe}]{sirseMeasurementCF3Densities2017}
Sirse, N., Tsutsumi, T., Sekine, M., Hori, M., \& Ellingboe, A.~R. 2017,
  Journal of Physics D: Applied Physics, 50, 335205,
  \dodoi{10.1088/1361-6463/aa77c4}

\bibitem[{Slater \& Michaelides(2019)}]{slaterSurfacePremeltingWater2019}
Slater, B., \& Michaelides, A. 2019, Nature Reviews Chemistry, 3, 172,
  \dodoi{10.1038/s41570-019-0080-8}

\bibitem[{Spilker {et~al.}(2003)Spilker, Ferrari, Cuzzi, Showalter, Pearl, \&
  Wallis}]{spilkerSaturnRingsThermal2003a}
Spilker, L., Ferrari, C., Cuzzi, J., {et~al.} 2003, Planetary and Space
  Science, 51, 929, \dodoi{10.1016/j.pss.2003.05.004}

\bibitem[{Stude {et~al.}(2021)Stude, Aufmhoff, Schlager, Rapp, Arnold, \&
  Strelnikov}]{studeNovelRocketborneIon2021}
Stude, J., Aufmhoff, H., Schlager, H., {et~al.} 2021, Atmospheric Measurement
  Techniques, 14, 983, \dodoi{10.5194/amt-14-983-2021}

\bibitem[{Sturm {et~al.}(2010)Sturm, Bouwman, Henning, Evans, Acke, Mulders,
  Waters, van Dishoeck, Meeus, Green, Augereau, Olofsson, Salyk, Najita,
  Herczeg, van Kempen, Kristensen, Dominik, Carr, Waelkens, Bergin, Blake,
  Brown, Chen, Cieza, Dunham, Glassgold, Gudel, Harvey, Hogerheijde, Jaffe,
  Joergensen, Kim, Knez, Lacy, Lee, Maret, Meijerink, Merín, Mundy,
  Pontoppidan, Visser, \& Yildiz}]{sturmFirstResultsHerschel2010}
Sturm, B., Bouwman, J., Henning, T., {et~al.} 2010, Astronomy and Astrophysics,
  518, L129, \dodoi{10.1051/0004-6361/201014674}

\bibitem[{Sugiyama(1994)}]{sugiyamaIonrecombinationNucleationGrowth1994}
Sugiyama, T. 1994, Journal of Geophysical Research, 99, 3915,
  \dodoi{10.1029/93JA02822}

\bibitem[{Terada {et~al.}(2007)Terada, Tokunaga, Kobayashi, Takato, Hayano, \&
  Takami}]{teradaDetectionWaterIce2007}
Terada, H., Tokunaga, A.~T., Kobayashi, N., {et~al.} 2007, The Astrophysical
  Journal, 667, 303, \dodoi{10.1086/520951}

\bibitem[{Tolman(1949)}]{tolmanEffectDropletSize1949}
Tolman, R.~C. 1949, The Journal of Chemical Physics, 17, 333,
  \dodoi{10.1063/1.1747247}

\bibitem[{Vahidinia {et~al.}(2011)Vahidinia, Cuzzi, Hedman, Draine, Clark,
  Roush, Filacchione, Nicholson, Brown, Buratti, \&
  Sotin}]{vahidiniaSaturnRingGrains2011}
Vahidinia, S., Cuzzi, J.~N., Hedman, M., {et~al.} 2011, Icarus, 215, 682,
  \dodoi{10.1016/j.icarus.2011.04.011}

\bibitem[{Vaste(1993)}]{vasteNoctilucentClouds1993}
Vaste, O. 1993, Journal of Atmospheric and Terrestrial Physics, 55, 133,
  \dodoi{10.1016/0021-9169(93)90118-I}

\bibitem[{Witt(1969)}]{wittNatureNocitlucentClouds1969}
Witt, G. 1969, Space Research, IX, 157

\bibitem[{Woitke {et~al.}(2009)Woitke, Kamp, \&
  Thi}]{woitkeRadiationThermochemicalModels2009}
Woitke, P., Kamp, I., \& Thi, W.-F. 2009, Astronomy \& Astrophysics, 501, 383,
  \dodoi{10.1051/0004-6361/200911821}

\bibitem[{Zasetsky {et~al.}(2009)Zasetsky, Petelina, \&
  Svishchev}]{zasetskyThermodynamicsHomogeneousNucleation2009a}
Zasetsky, A.~Y., Petelina, S.~V., \& Svishchev, I.~M. 2009, Atmospheric
  Chemistry and Physics, 9, 965, \dodoi{10.5194/acp-9-965-2009}

\end{thebibliography}

\end{document}